\documentclass[letterpaper, 10 pt, conference]{ieeeconf}  

\IEEEoverridecommandlockouts                              

\usepackage{import}
\usepackage{graphics} 
\usepackage{epsfig} 
\usepackage{algpseudocode}
\usepackage[linesnumbered,ruled,vlined]{algorithm2e}
\usepackage{times} 
\usepackage{amsmath} 
\usepackage{amssymb}  
\usepackage{siunitx}
\usepackage{mathtools}
\usepackage{xfrac}
\usepackage{hyperref}
\usepackage{multirow}
\usepackage{array}
\title{\LARGE \bf Signal Quality Assessment of Photoplethysmogram Signals using Quantum Pattern Recognition and lightweight CNN Architecture}

\author{Tamaghno Chatterjee, Aayushman Ghosh and Sayan Sarkar  
\thanks{Draft Version. Currently submitted at IEEE EMBC 2022}
\thanks{*This work was supported partially by EUREKA’16 start-up grant.}
\thanks{Aayushman Ghosh and Tamaghno Chatterjee are with Indian Institute of Engineering Science and Technology, Shibpur. Sayan sarkar is associated with Hong Kong University of Science and Technology.
         {\tt\small ssarkar@connect.ust.hk}}}%

\begin{document}

\maketitle
\thispagestyle{empty}
\pagestyle{empty}


\begin{abstract}
Photoplethysmography (PPG) signal comprises physiological information related to cardiorespiratory health. However, while recording, these PPG signals are easily corrupted by motion artifacts and body movements, leading to noise enriched, poor quality signals. Therefore ensuring high-quality signals is necessary to extract cardiorespiratory information accurately. Although there exists several rule-based and Machine-Learning (ML) - based approaches for PPG signal quality estimation, those algorithms' efficacy is questionable. Thus, this work proposes a lightweight CNN architecture for signal quality assessment employing a novel Quantum pattern recognition (QPR) technique. The proposed algorithm is validated on manually annotated data obtained from the University of Queensland database. A total of 28366, 5s signal segments are preprocessed and transformed into image files of 20 × 500 pixels. The image files are treated as an input to the 2D CNN architecture. The developed model classifies the PPG signal as `good' or `bad' with an accuracy of 98.3\% with 99.3\% sensitivity, 94.5\% specificity and 98.9\% F1-score. Finally, the performance of the proposed framework is validated against the noisy `Welltory app' collected PPG database. Even in a noisy environment, the proposed architecture proved its competence. Experimental analysis concludes that a slim architecture along with a novel Spatio-temporal pattern recognition technique improve the system's performance. Hence, the proposed approach can be useful to classify good and bad PPG signals for a resource-constrained wearable implementation.


\end{abstract}
\section{INTRODUCTION}

PPG is an essential physiological signal representing arterial oxygenation versus time. It carries useful physiological information like heart rate (HR), blood pressure (BP), blood oxygen saturation levels, respiration rate (RR), arterial stiffness, vascular ageing etc. [1-30]. PPG signals are sensitive to multiple biological, environmental factors that impact measurement results [2]. A slight movement of the arm, poor blood perfusion of the peripheral tissues, or introduction of ambient light at the photodetector can significantly corrupt the PPG signals' morphology [3-4]. These distortions are more apparent in PPG signals recorded via mobile devices, which are often corrupted by noise and artifacts that subsequently lead to a significant number of misleading diagnoses [1-3],[29]. An example of how corrupted PPG segments can influence the extraction of physiological parameters is shown in Fig.1 (a). The maximum frequency in the clean PPG spectrum is 1.1 Hz, which gives 66 bpm in HR estimation. However, the PPG signal in Fig.1 (b) is corrupted by motion artifacts. The maximum frequency is 0.19 Hz in the PPG spectrum leading to an HR estimation of 12 bpm, which is significant enough to trigger a false alarm. Moreover, the presence of artifacts causes alarm fatigue, signal loss, and inaccurate measurements and diagnoses [2-15]. Therefore, to identify distorted PPG segments, Elgendi [2] proposed eight rule-based signal quality indexes (SQI) like perfusion, kurtosis, skewness, etc. The 'skewness index' outperformed the other seven indices in good and bad signals differentiation [2]. J. A. Sukor et al. [5] proposed an auto-rejection method for artifact-contaminated PPG waveforms using the pulse wave analysis (PWA) technique and a decision tree classifier. Q Li et al. [6] introduced a dynamic time warping (DTW) method to stretch the single-cycle pulse waveform for a running template. This method, combined with several other signal quality features are  fed to a multilayer perceptron (MLP) neural network to estimate signal quality. R Couceiro et al. [7] proposed a motion artifact detection algorithm based on the variation of PPG signals' time-domain features. They used the SVM classifier for distinguishing clean and corrupted signals. S. Cherif et al. [8] described a new method based on PWA to detect PPG signal artifacts using a random distortion test relied on 'adaptive thresholding'.
\vspace{-1.5mm}
\begin{figure}[h]
    \centering
    \includegraphics[width = 8.6 cm, height = 6cm]{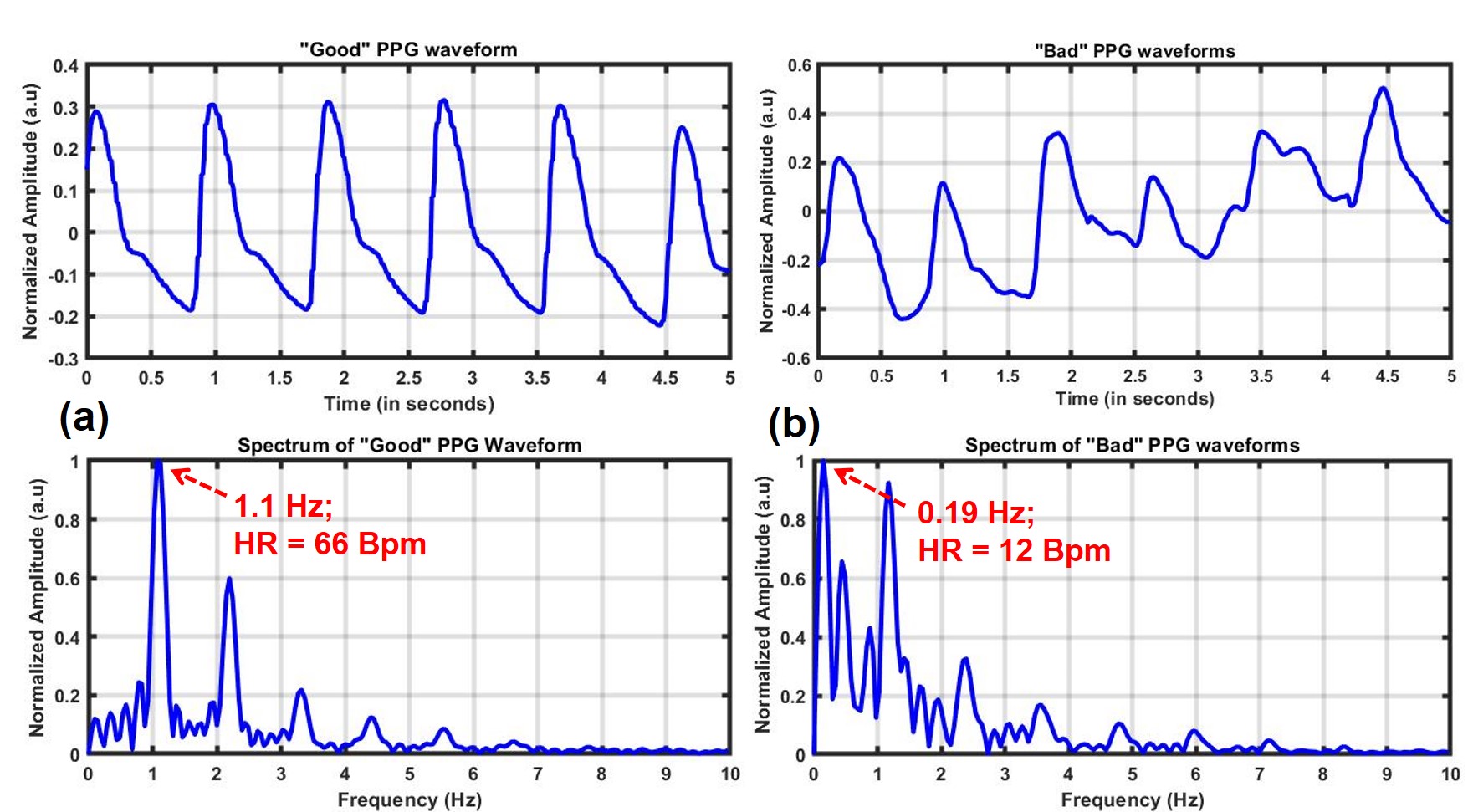}
    \caption{PPG signals with their spectrum (a) good signal, (b) bad signal}
    \label{Problems in PPG signals for cardiovascular parameter estimation}
\end{figure}
\vspace{-1.5mm}
Multiple studies have proposed ML [9-11] and DL [12-17] architectures for the signal quality assessment (SQA) by considering the above rule-based methods' accuracy limitation. DL has gained widespread attention due to its powerful ability to automatically learn from physiological signals, such as electrocardiograms (ECGs) and PPGs. Prominent DL architecture Convolutional Neural network (CNN) is widely used in the ECG classification, BP, HR estimation and biometric identification from PPG signals [12- 17]. S.-H. Liu et al.[9] evaluated the qualities of PPG signals with CNN and support vector-based approaches and achieved lower error ratios of oxygen saturation ratio (SpO2) after classification. Later, Liu et al. [11] have proposed a five-layer fuzzy neural network-based SQA algorithm to assess the signal quality of PPG and reported a sensitivity below 0.9. C. H. Goh et al. [12] designed a 1D Convolutional Neural Networks (CNN) model to classify five-second PPG segments into a clean or artifact-affected segment. In another study, Liu et al. [16] have classified PPG quality into three levels (high, middle, and low) using PPG and differential PPG as inputs for a 2D deep convolution neural network and residual deep CNN. Recently, both the time and frequency domain information [15] have been used for signal quality estimation using short term Fourier transform (STFT) spectrum. 1D PPG signals are first transformed into 2D STFT images and the 2D CNN model takes those images as the inputs and performs the classification task [15]. Although traditional neural network-based models can achieve high PPG classification accuracy, they are often parameter intensive and increase the whole system's computational complexity[12-17]. Thus they are not suitable for deployment in resource-constrained computing platforms. The authors propose a lightweight 2D CNN model based on novel Slim module micro-architecture to address this problem. The 1D PPG signals are first converted to grayscale images using a novel Quantum Pattern Recognition algorithm and these images are classified into 'Good' and 'Bad' categories using the proposed CNN model.
\vspace{0.5mm}
\section{MATERIALS AND METHODS}

\subsection{Database Description} 
The University of Queensland vital signs dataset (accessed on July 2021) was used in this study [18]. This is an open-access public database of intra-operative vital signs and biosignals collected at the `Royal Adelaide Hospital' [18]. The dataset was recorded from 32 subjects using Phillips IntelliVue MP70 and MP30 devices at a sampling rate of 100 Hz. The database contains signal recordings from each patient, ranging between 13 minutes to 5 hours [18]. Multiple groups used this database for BP classification and estimation purposes [19]. We extracted signal segments of length 5s (500 samples) from all the recordings and manually annotated those segments as 'Good' or 'Bad' with the help of expert annotators. Detailed information regarding the number of segments in each category for all 32 cases is provided in Table I. Authors have chosen another PPG database [20] (accessed on Nov 2021 - collected with smartphones using the Welltory app) to test the proposed algorithm on artifact enriched PPG waveforms. The PPG signals were recorded at a sampling rate of 125 Hz. This publicly available dataset consists of recordings from 21 subjects. Each record contains three-time series PPG signals (red, green and blue channel), among which the red channel data is available for all the 21 recordings [20]. Hence authors have considered only red channel data in this paper. 4s of signal segments (500 samples) from each of the 21 recordings were extracted to keep sync. with the length of the signals extracted from the Queensland dataset. Altogether, 94 signal segments were obtained.
\vspace{-1.5mm}
\begin{table}[h]
    \caption{ANNOTATED SEGMENTS IN THE QUEENSLAND DATABASE}
    \centering
    \setlength\tabcolsep{3.2 pt}
    \setlength{\extrarowheight}{0.5 pt}
    \begin{tabular}{l| cccc|c}
    \hline\hline
        Case & Good & Bad & Without & Un-categ- & Total \\
        (Subjects) & signals & signals & reference BP & orized & \\
        \hline
         Case 1 & 1131 & 220 & 67 & 0 & 1418\\
         Case 2 & 46 & 57 & 63 & 32 & 198\\
         Case 3 & 2086 & 360 & 162 & 26 & 2634\\
         Case 4 & 1339 & 225 & 123 & 0 & 1687\\
         Case 5 & 655 & 170 & 117 & 0 & 942\\
         Case 6 & 749 & 110 & 90 & 91 & 1040\\
         Case 7 & 402 & 90 & 166 & 133 & 791\\
         Case 8 & 414 & 216 & 41 & 0 & 671\\
         Case 9 & 745 & 216 & 89 & 218 & 1268\\
         Case 10 & 279 & 175 & 27 & 13 & 494\\
         Case 11 & 839 & 146 & 60 & 0 & 1045\\
         Case 12 & 958 & 195 & 193 & 36 & 1382\\
         Case 13 & 832 & 270 & 41 & 0 & 1143\\
         Case 14 & 695 & 230 & 101 & 0 & 1026\\
         Case 15 & 116 & 128 & 41 & 21 & 306\\
         Case 16 & 106 & 216 & 294 & 1571 & 2187\\
         Case 17 & 146 & 60 & 25 & 0 & 231\\
         Case 18 & 96 & 24 & 14 & 0 & 134\\
         Case 19 & 202 & 35 & 27 & 0 & 264\\
         Case 20 & 995 & 175 & 104 & 0 & 1274\\
         Case 21 & 583 & 100 & 137 & 0 & 820\\
         Case 22 & 404 & 111 & 152 & 21 & 688\\
         Case 23 & 220 & 96 & 12 & 0 & 328\\
         Case 24 & 557 & 135 & 29 & 0 & 721\\
         Case 25 & 783 & 135 & 206 & 0 & 1124\\
         Case 26 & 1418 & 270 & 271 & 0 & 1959\\
         Case 27 & 1009 & 380 & 246 & 563 & 2198\\
         Case 28 & 520 & 253 & 52 & 42 & 867\\
         Case 29 & 1009 & 306 & 92 & 0 & 1407\\
         Case 30 & 517 & 95 & 93 & 0 & 705\\
         Case 31 & 2058 & 380 & 41 & 0 & 2479\\
         Case 32 & 728 & 150 & 72 & 0 & 950\\
         \hline
         \end{tabular}
    \label{tab:my_label1}
\end{table}
\vspace{-3mm}
\subsection {Signal quality determination}

\begin{figure}[h]
    \centering
    \includegraphics[width = 8.6 cm, height = 5.1 cm]{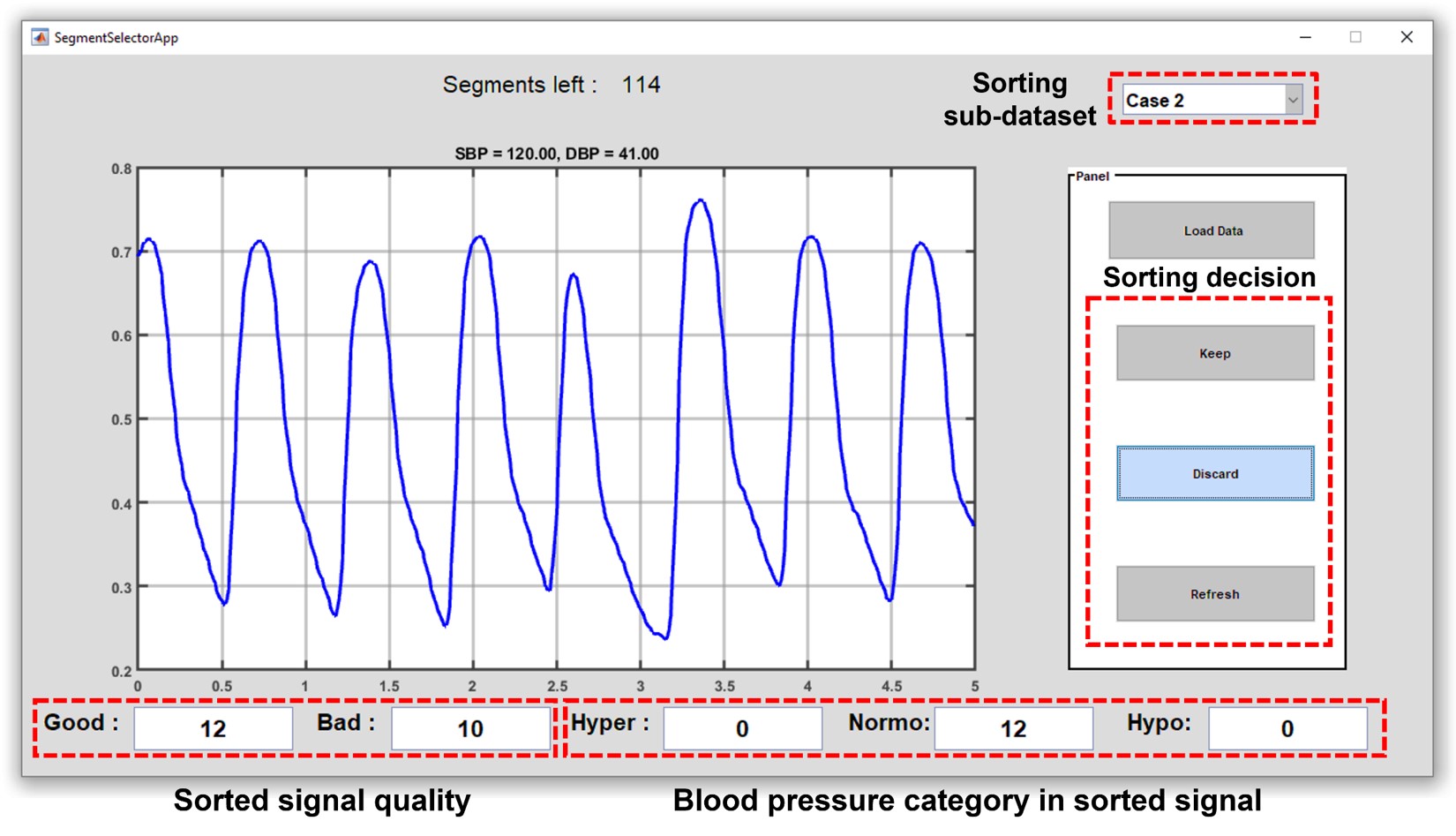}
    \caption{GUI platform for signal quality classification}
    \label{GUI platform for signal quality classification}
\end{figure}

From each case file of the Queensland dataset, raw PPG signal waveform and their corresponding discrete BP data (SBP, DBP, MAP) were extracted. A 5-second long non-overlapping window was slid over each extracted data file to generate PPG segments. Each PPG segment was annotated either in the ‘Good’, or ‘Bad’ category by expert annotators with the help of a GUI shown in Fig. \ref{GUI platform for signal quality classification} [5-13]. A good PPG signal is defined as: (1) the PPG signal has a clear and undisturbed waveform, (2) the reflection points of the waveform are relatively consistent [2]. Although the intended use case in this study is SQI but eventually, as a part of future works, this dataset will be explored for BP estimation or classification purposes. Hence, the developed GUI platform in Fig. \ref{GUI platform for signal quality classification} also involves the BP categorization details. Before the data segmentation, the PPG segments without corresponding reference BP data are discarded. The manual check was also necessary to identify and exclude those PPG segments that fall in the `un-categorized' section where the reported BP values fall below the standard hypotensive category [18]. Hence those PPG segments were not considered while performing further analysis. Earlier, Khalid et. al [19] segregated the `Queensland database' into multiple segments. However, a detailed manual observation reveals discrepancies in the sub-groups. Sometimes the total number of signals is reduced without any description, or the number of entries in each sub-group does not match with the raw dataset [19]. These discrepancies pushed the authors to reevaluate the whole scenario starting from scratch. A total of 28366 segments of signals were selected, out of which 22637 segments are good-quality signals and 5729 segments are bad quality signals as per TABLE I. A similar method was followed to manually annotate the `Welltory' dataset that yielded 52 `Good' and 42 `Bad' segments.
\vspace{-2.5 mm}
\vspace{-3.5 mm}
\subsection{Quantum based spatio-temporal pattern recognition}

Several studies have used 2D CNN [13-14] for SQI based PPG classification owing to it's success in image classification tasks. However, 2D CNNs require image data, whereas PPG signals are one dimensional in nature [4-5]. Therefore, authors propose a novel quantum-based pattern recognition technique that accurately learns the Spatio-temporal features of a 1D data, and map them to two-dimensional images. This model is originally inspired from the Semi-Classical Signal Analysis (SCSA) technique proposed by Laleg \textit{et. al} [22], in which an input signal is treated as an attractive potential within the Schrödinger operator. From the discrete spectrum of the Schrödinger operator, the input signal is accurately reconstructed. SCSA method has previously been applied to cuff-less BP estimation [23], Magnetic Resonance Spectroscopy (MRS) denoising [24] and spike artifact removal from EEG-fMRI data [25]. Here, authors' have assumed a real-valued, time-dependent non-negative input signal $y(t)$ and a positive parameter $h$, popularly known as the `semi-classical parameter' as the input arguments. The 1D semi-classical time-dependent Schrödinger operator $\mathcal{H}_{h}(y,t)$ follows (1).
\vspace{-1.5 mm}
\begin{equation}
\mathcal{H}_{h}(y,t) = -h^{2}\frac{d^{2}}{dt^{2}} - y(t),\quad t \in\mathbb{R}, h \in \mathbb{R}_{+}^{*},
\label{equation1}
\end{equation}
Such that 
\begin{equation}
\mathcal{H}_{h}(y,t)\psi(t) = \lambda\psi(t)
\label{equation2}
\end{equation}
The Schr\"odinger equation represented as an eigenvalue problem follows (3).
\begin{equation}
-h^{2} \frac{d^{2} \psi(t)}{d t^{2}}-y(t) \psi(t)=\lambda \psi(t)
\end{equation}
Where $\lambda$ is the eigenvalue and $\psi(t)$ is the corresponding $L^2$ normalized eigenfunction on which $\mathcal{H}_{h}(y,t)$ operates [22]. In general, there exists a continuous spectrum for $\lambda \geq 0$ and a discrete spectrum for $\lambda < 0$ such that
\begin{equation}
    \lambda_{nh} = -{\kappa}_{nh}^2, \quad \int_{-\infty}^{+\infty}\psi_{nh}^2 = 1, \forall n= 1,2,\cdots,N_h
\end{equation}
where $\lambda_{nh}$ represents the negative eigenvalues with $\kappa_{nh}>0$, and ${\lambda_{1h}\leq\lambda_{2h}\cdots\lambda_{N_{h}h}<0}$. $N_h$ is defined as the number of negative eigenvalues. The SCSA technique uses this negative eigenvalues and its corresponding eigenfunctions to reconstruct the original signal $y(t)$. The reconstructed signal $y_h(t)$ follows (5) [23-25]. 
\begin{equation}
y_{h}(t)=4 h \sum_{n=1}^{N_{h}} \kappa_{n h} \psi_{nh}^{2}(t), t \in \mathbb{R}
\end{equation} 
Based on discrete spectrum of the operator $\mathcal{H}_{h}(y,t)$, the input is decomposed into a series of smooth pulse-shaped orthogonal basis functions called Schrödinger components ($4h\kappa_{1h}\psi_{1h}^{2}(t)$, $4h\kappa_{2h}\psi_{2h}^{2}(t)$, $4h\kappa_{3h}\psi_{3h}^{2}(t)$, $\cdots$ $\forall$ $n\in 1,2,3,\cdots,N_h$), as shown in Fig. 3. These basis functions are localized and selectively capture the Spatio-temporal pattern of $y(t)$.
\vspace{-1.5mm}
\begin{algorithm}
    \caption{Function: SCSA Reconstruction}
    \SetAlgoLined
    \SetKwInOut{Input}{Input}
    \SetKwInOut{Output}{Output}
    \SetKwRepeat{Repeat}{Repeat}
    \DontPrintSemicolon
    \SetKwFunction{FMain}{SCSAReconstruction}
    \SetKwProg{Fn}{Function}{:}{}
    \Fn{\FMain{$h, y(t), N_h$}}{
    \textbf{Begin:} Discretize $2^{nd}$-order Differentiation matrix $\mathcal{D}_2$, using $fs$ \& $max(size(y(t)))$,
    
    Diagonalize $y(t)$ \& using $\mathcal{D}_2$, construct $\mathcal{H}_h(y,t)$ 
    
    Solve (3), to extract $\lambda$, $\psi$
    
    $\lambda_{nh} = find(\lambda<0)$, $\kappa_{nh} = diag( h\sqrt{-\lambda_{nh}})$
    
    $\psi_{nh}$ = $\psi$(position of $\lambda_{nh}$), normalize over $L^2$
    
    $\mathcal{P} = 4h\kappa_{nh}\psi_{nh}^{2}$
    
    \If{$size(\kappa_{nh}) \geq N_h$}{
    
    $y_{h}$ = $sum(\mathcal{P}(1:N_h, :))$
    \%{row wise sum}
    
    $\mathcal{O}$ = $\mathcal{P}(1:N_h, :)$ \%{Sch. Components}
    }
    \textbf{return} [$\mathcal{O}, y_{h}$]
    }
    \textbf{end Function}
    \label{Algorithm_1}
\end{algorithm}
\vspace{-1.5mm}
As evident from (5), Schrödinger components are squared eigenfunctions scaled by their respective eigenvalues [22], making SCSA analogous to the Fourier transform, which uses complex sinusoids to decompose an input [15], [25]. Algorithm 1 extracts these basis functions $\mathcal{O}$, along with the reconstructed signal ${y_{h}}$. Low-order Schrödinger components capture energy-dense peaks of the input, making this technique well-suited for peak retention applications or in broad-scale pattern recognition. It has been previously shown [22-25] that the parameter $h$ plays a critical role in SCSA based reconstruction of the signal $y(t)$. For a large value of $h$, fewer components are formed in the series expansion of (5). These components fail to capture the precise details of $y(t)$, implementing a bad reconstruction [25]. It can be observed that, by lowering the value of $h$ ($h\xrightarrow{}0$), higher number of Schr\"odinger components are formed. These higher-order components capture the input signal's finer details $y(t)$, facilitating high-fidelity reconstruction. Proposed Quantum Pattern Recognition Algorithm is based on these observations. Authors observed that at least 20 level decomposition of PPG signal segments is necessary to fully capture the morphological variations of the input signal, achieving minimum reconstruction error. The signal reconstruction in Fig. 4(b) affirms excellent spatio-temporal pattern recognition. The decomposition levels (here 20 levels) generated in the process is used to construct a gray-scale image from the 1D input signal as described in the processing pipeline. This image is used as an alternative form of representation for the PPG signal analogous to the STFT transform in Fig. 4(c).
\vspace{-1.5mm}
\begin{figure}[h]
    \centering
    \includegraphics[width=8.6cm, height = 5.05cm]{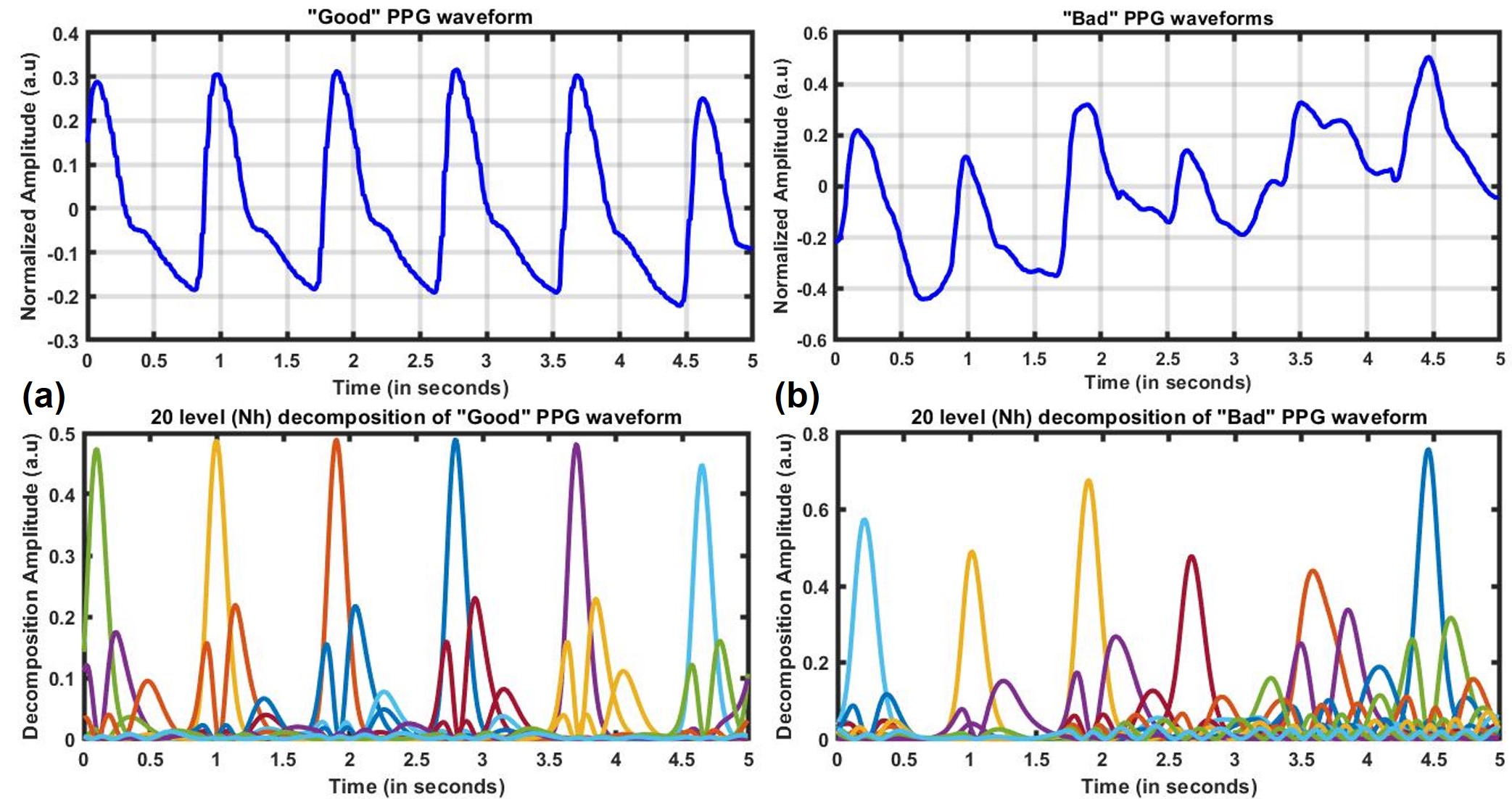}
    \caption{Decomposition of a 5 seconds PPG signal segment into pulse-shaped Schr\"odinger Components (Here 20 level decomposition)}
    \label{Estimated systolic and diastolic dynamics. PPG signal segment (in red) and SCSA estimated parts (in dotted lines).}
\end{figure}
\vspace{-3.5mm}
\begin{algorithm}
    \caption{Quantum Pattern Recognition}
    \SetAlgoLined
    \SetKwInOut{Input}{Input}
    \SetKwInOut{Output}{Output}
    \SetKwRepeat{Repeat}{Repeat}
    \DontPrintSemicolon
    \Input{$y(t) \gets$Time-dependent positive Input signal, \linebreak
    $N_h \gets$ No. of negative eigenvalue, 
    \linebreak
    $[h_{min}, h_{max}] \gets$ Suitable $h$ range 
    }
    \Output{Image Matrix: $\mathcal{I}_h$}
    $\Omega_k = linspace(h_{min}, h_{max}, floor(\frac{N_h}{2}))$
    
    
    $\epsilon \gets$ \texttt{Vector}(size: $length(\Omega_k)$)

    \For{$i \xrightarrow{} 1:length(\Omega_k)$ }{
    $h = 1/(\Omega_k(i))^2$
    
    [$\sim, y_{h}$] = \texttt{SCSAReconstruction}($h, y(t), N_h$)
    
    \If{$numel(y_{h}>0)$ AND $sum(y_{h}$ != $0)$}{
    $\epsilon(i)$ = $||y - y_{h}||$
    }
    }
    
    $\Omega_k = \Omega_k(\epsilon>0)$, $\mathcal{E} = \epsilon(\epsilon>0)$
    
    \If{$numel(\Omega_k)>0$}{
    
    $\chi = min(\mathcal{E})$, $\hat{h} = {1}/{(\Omega_k(\chi))^2}$
    
    [$\mathcal{O}, \hat{y_{h}}$] = \texttt{SCSAReconstruction}($\hat{h}, y(t), N_h$)
    }
    $\mathcal{I}_h \gets$ $\mathcal{O}./max(|{\mathcal{O}}|)$ \%\textbf{Matrix}($N_h$ $\times$ $length(y(t))$
    \label{Algorithm_2}
\end{algorithm}
\vspace{-1.5mm}
\section{PROCESSING PIPELINE}

The processing pipeline was designed to convert amplitude normalized 1D PPG signals into 2D images by following algorithms 1 and 2. A 20 level decomposition of each $1\times500$ PPG segment is performed using Algorithm 1. The 20 decomposition levels (each of $1\times500$ dimension) are concatenated vertically to obtain a $20\times500$ matrix. Finally, each value in the matrix was divided by the maximum value of that matrix. Each of the matrices can be considered as a $20\times500$ pixel grey-scale image in Fig. 4(d) which is used as an input to the proposed CNN architecture.
\vspace{-3.5mm}
\subsection{Deep learning architecture}
A human being can differentiate between two images by carefully observing and remembering the small distinguishing features. A 2D CNN model does the same thing by extracting features in each layer and from these features, it learns to distinguish between images. This paper has discussed a computationally-efficient 2D CNN architecture based on slim Module micro-architecture. The Slim-CNN architecture was previously used in face attribute prediction [26]. Earlier studies [16-17], have explored the ResNet or Residual Network architecture in the domain of SQI based PPG classification. ResNet and slim-CNN both utilise multiple branches for a better representation of discriminative features [26]. However, the slim-CNN efficiently uses depthwise separable convolutions with pointwise convolutions [26-29] in its design. This approach results in significant reduction in the number of trainable parameters and on-disk memory while maintaining accuracy comparable to ResNet [26]. PPG signal's intended use case is a resource-constrained computing device like a mobile phone or smartwatch, therefore, the novel slim-CNN-based architecture was preferred over parameter intensive ResNet architecture. The proposed CNN architecture for PPG based SQA follows Fig.5. The network's input is the $20\times500$ pixel image derived from PPG segments using the quantum pattern recognition algorithm. A $7\times7$ 2D convolution layer combined with Batch normalization and ReLU activation function is placed right next to the input layer. A $2\times2$ Maxpool layer with stride = $2$ was used to reduce number of pixels. The output of the Maxpool layer is fed to the Slim Module section. In the proposed architecture, three Slim Modules are arranged in cascade, with each having a skip connection between its input and the output node. The skip connects reduce the information attenuation and provide flatter loss surfaces during training [26], thus helping the model converge while getting rid of local minimas. $2\times2$ Maxpool layers with stride = $2$ were placed between the Slim Modules as per Fig. 5. The output of the Slim Module section is fed to the Global Average Pooling (GAP) layer since it is superior both in terms of computational efficiency and performance compared to dense layers [26-29]. The output of the GAP layer is passed through a dense layer of size 64 with ReLU activation and a dropout layer with a dropout rate of 0.5 before finally passing it through a sigmoid classifier. The filter counts and structure of the Slim Modules were kept the same as of [26]. All the convolution layer kernels were initialized with Xavier's method [26-29], and the dense layer before the sigmoid layer was only initialized with Kaiming's method [28-29].
\vspace{-1mm}

\begin{figure*}[h]
    \centering
    \includegraphics[width = 17.6 cm, height = 5.7cm]{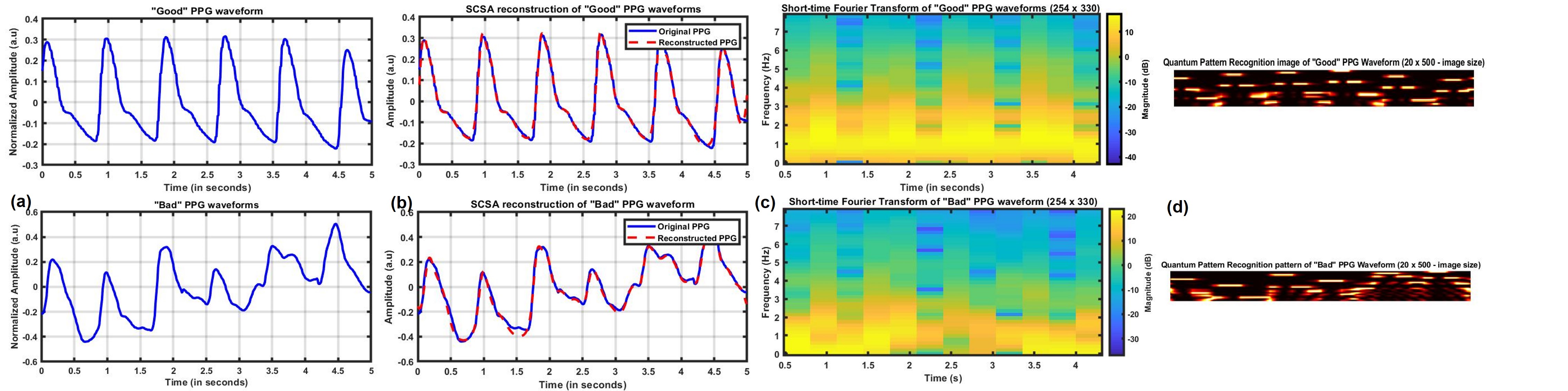}
    \caption{Various representations of "Good" and "Bad" PPG signals (a) Extracted PPG signal segments (b) SCSA based reconstructed waveforms (red dotted line) (c) STFT transform of the PPG segments, (d) QPR-based generated image of the PPG signal segment}
    \label{Different PPG signal quality and scsa based operation}
\end{figure*}
\begin{figure*}[h]
    \centering
    \includegraphics[width = 17.5 cm, height = 7.6cm]{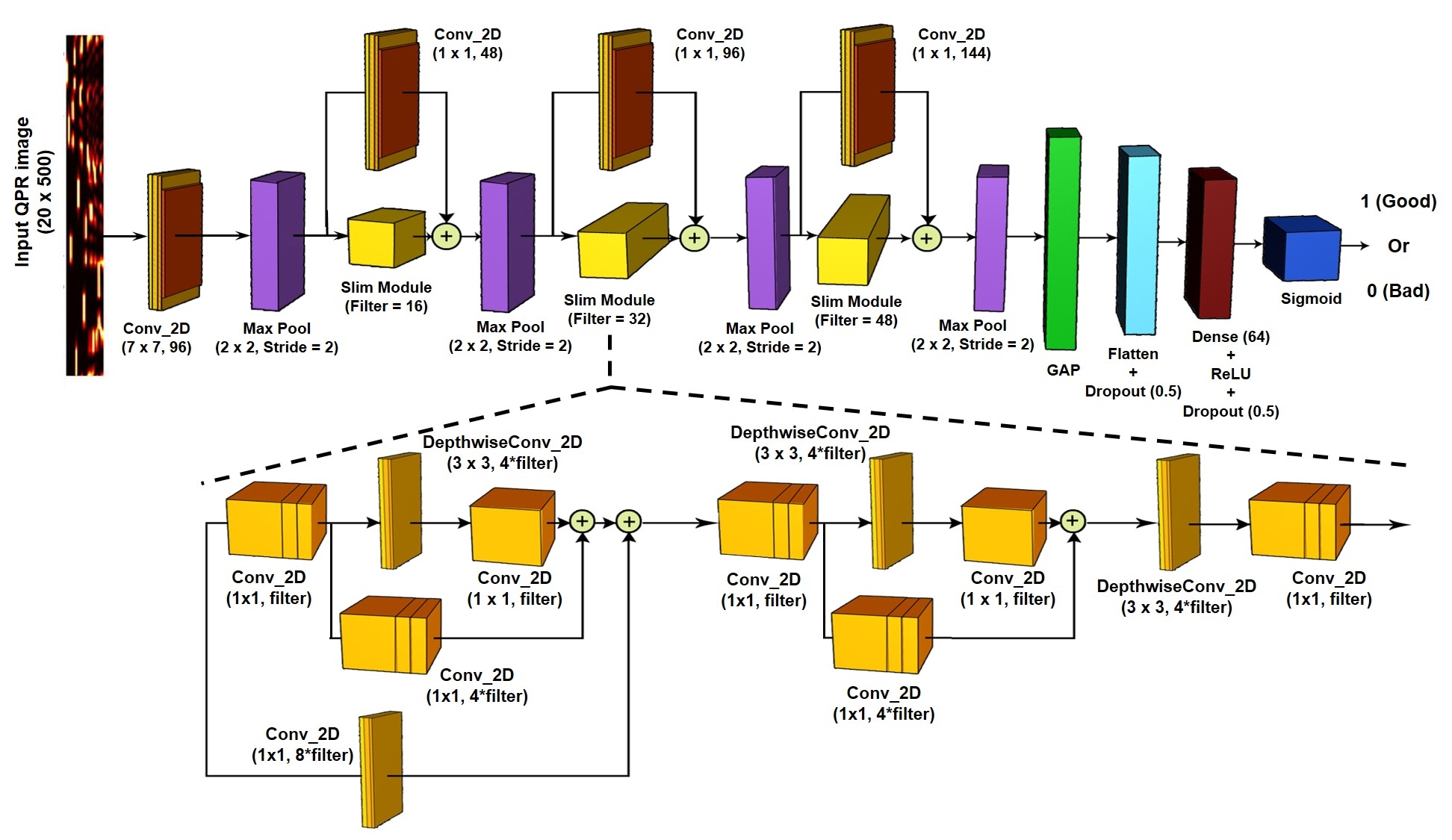}
    \caption{Proposed lightweight CNN architecture for QPR Image processing. (Conv$\_$2D = Convolution 2D + Batch Normalization + Rectified Linear Unit (ReLU); DepthwiseConv$\_$2D = Depthwise Convolution 2D + Batch Normalization + Rectified Linear Unit (ReLU)); GAP = Global Average Pooling).}
    \label{Proposed Slim-CNN architecture for image processing}
\end{figure*}
\subsection {Simulation environment}

The model is built using the Keras API of the Tensorflow 2.7.0 platform and Python 3.7.12 for programming in the 'Jupyter notebook' environment [15]. Finally, the model runs on an Intel(R) CPU @2.3 GHz computer with 32 GB DDR4 memory, with a windows 11 operating system. The graphics card is NVIDIA® GeForce RTX™ 3060 with 6 GB dedicated memory for acceleration. The model uses the Adaptive Moment Estimation (Adam) optimizer during training and binary cross-entropy as the loss function. The training epochs are 80, and the batch size is 80, with a learning rate of 0.005 [15].

\section{STATISTICAL ANALYSIS}

\subsection{Evaluation metrics}

Different metrics exist in the literature to evaluate a classifier's performance. These include: Accuracy, sensitivity, specificity and F1 scores [5-15]. To calculate these, True Positives(TP), True Negatives(TN), False Positives(FP) and False Negatives(FN) [5-15] are obtained from the confusion matrix. The receiver operating characteristic (ROC) curves and the area under the receiver operating characteristic curve (AUC) are also used in literatures [5-15].




\subsection{Test results of the proposed method}
The Queensland dataset was used to evaluate the classification model's performance on a large number of signal segments. A total of 28366 images each of size $20\times500$ pixels and corresponding one-hot encoded labels were used to generate Train and Test data. The training set was constructed using a random sample of 22692 (80\%) images and the rest (5674, 20\%) were used to construct the test set. 10\% of the training data was used for validation during training. Test results based on accuracy, sensitivity, specificity, F1-score and support values are reported in Table  II. Fig. 6(a) shows the confusion matrix of the proposed model. Out of the 5674 image segments, 5575 are classified correctly, and 99 are misclassified. The Receiver Operating Characteristic curve (ROC) and the AUC score is presented in Fig. 6 (a) and (b). The AUC of 0.992 presents outstanding discrimination performance [15]. Post evaluation of model's performance on Queensland dataset, authors evaluated the algorithm's efficacy on noisy Welltory dataset. Since, the Welltory dataset contains recordings from a smartphone device and recording was done in a non-clinical set-up, the signals contain significant motion artifact and noise. Although the size of the welltory database is short but it is comparable to earlier work [8]. The images obtained from 94 signal segments were tested on the model trained previously using the Queensland dataset. The results are presented in Table II. An estimation accuracy of 96.7\% with an AUC of 0.969 establishes the model's generalization ability along with the QPR algorithm's noise immunity.
\vspace{-2mm}
\begin{figure}[h]
    \centering
    \includegraphics[width = 8.65 cm, height = 4.3cm]{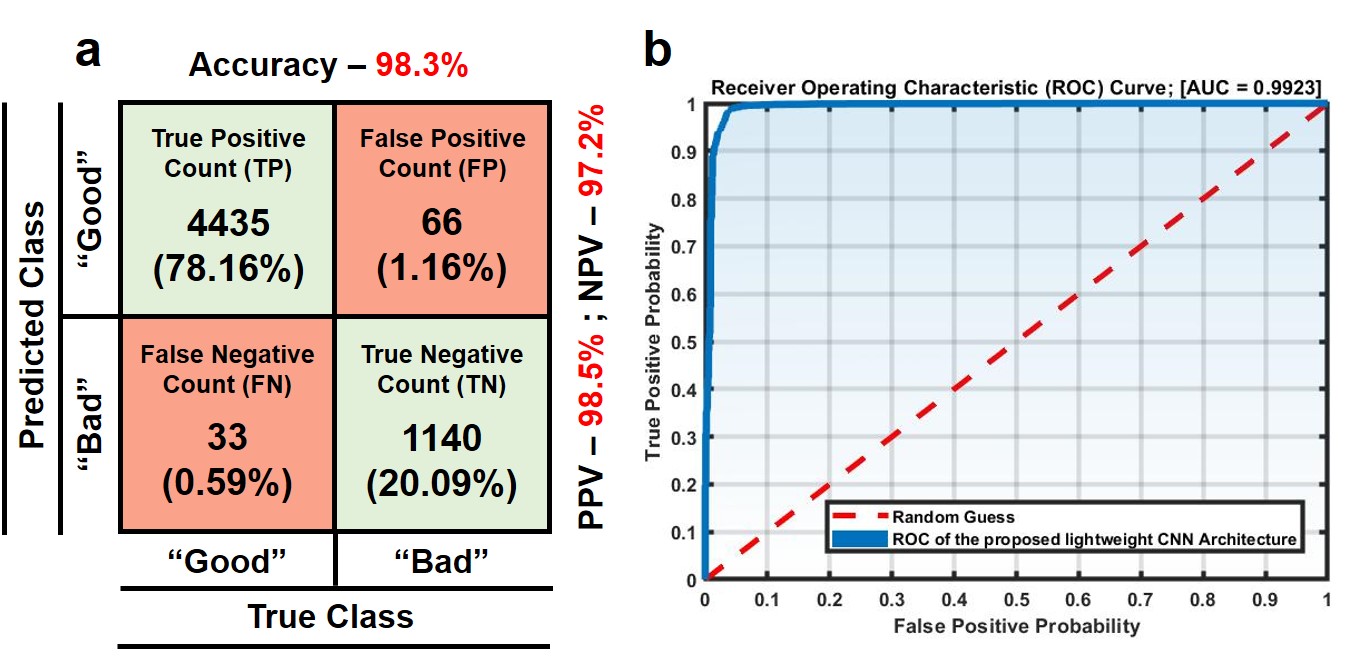}
    \caption{(a) Confusion Matrix with the corresponding Predictive Value (PPV) and Negative Predictive Value. (b) ROC curve for University of Queensland database}
    \label{The ROC curve }
\end{figure}
\vspace{-3.5mm}
\begin{table}[h]
    \caption{EVALUATION RESULTS}
    \centering
    \setlength\tabcolsep{3.2 pt}
    \begin{tabular}{c cccccc}
    \hline\hline
        Dataset (Train-Test) & Se & Sp & F1-score & ACC & Support \\
        \hline
         Queensland-Queensland & 99.3\% & 94.5\% & 98.9\% & 98.3\% & 5674 \\
         Queensland-Welltory & 100\% & 92.5\% & 97.2\% & 96.7\% & 94 \\
     \hline
    \end{tabular}
    \label{tab:my_label2}
\end{table}
\vspace{-3mm}
\subsection {Compare with established baseline models}
A comparative analysis is essential to validate the effectiveness of the proposed method. Three traditional methods are selected as baseline models and evaluated against the Queensland dataset using the proposed algorithm. These baseline models are (i) multiple SQIs-SVM, (ii) HOG-MLP, (iii) 1D CNN. Multiple studies [2],[4] have used standalone SQIs to estimate empirical thresholds for the classification of PPG signals. The performance of standalone SQIs is not upto the mark for classification jobs since they give a high number of false positives, resulting in low specificity [5]. So, in this paper, multiple SQIs calculated from each signal segment were combined together to form input feature set for a non-linear SVM classifier utilising the 'rbf' kernel. Histogram of Oriented Gradients (HOG) is a popular algorithm that extracts vital features from pedestrian trajectories or ECG images [15]. These features are then used by ML or DL models to classify images. A total of 1764 HOG features similar to earlier work [15] were extracted from the QPR images of PPG using the scikit-image library in Python. These features are then used as the input to a Multi-layer Perceptron Classifier. The 1D CNNs are chosen as a baseline method for DL based SQI estimation as it can extract morphological features from 1D time-series data. Whole PPG segments were used as inputs to the model. The results of the comparative analysis is presented in Table III. In terms of the four evaluation metrics in Table III, the proposed model outperformed all the baseline models.

\begin{table}[h]
    \caption{PERFORMANCE COMPARISON WITH DIFFERENT STANDARD METHODS}
    \centering
    \setlength\tabcolsep{3.2 pt}
    \begin{tabular}{l cccccc}
    \hline\hline
        Methods & Se & Sp & F1-score & ACC  \\
        \hline 
         SQIs+SVM & 90.6\% & 59.5\% & 90.3\% & 84.4\% \\
         \hline 
         1D-CNN & 97.7\% & 88.4\% & 97.4\% & 95.8\% \\ 
         \hline
         MLP+HOG & 95.7\% & 44.1\% & 91\% & 85\% \\
         \hline
         Proposed & \textbf{99.3\%} & \textbf{94.5\%} & \textbf{98.9\%} & \textbf{98.3\%}  \\
          \hline
    \end{tabular}

    \label{tab:my_label3}
\end{table}
\vspace{-3mm}
\subsection{Comparison with other available architectures}
The number of subjects and the size of the dataset must be considered for fair comparison since a model's generalisation ability is established when it is evaluated on large data. So, it is essential to evaluate the performance of two different CNN architectures on the same dataset. As the codes of [15] is open-sourced, authors reevaluated the work with the Queensland database first with proposed QPR based images, then with STFT images. In all evaluation categories, the performance of the proposed architecture outperformed the earlier DL based work as per Table IV. 

\begin{table}[h]
    \caption{PERFORMANCE COMPARISON WITH DL BASED METHOD IN QUEENSLAND DATASET}
    \centering
    \setlength\tabcolsep{2.8 pt}
    \setlength{\extrarowheight}{1pt}
    \begin{tabular}{l ccccc}
    \hline\hline
        & \multicolumn{5}{l}{USING QUEENSLAND DATASET - QPR images}\\
         \hline
         Dataset & Author & Se & Sp & ACC & F1-Score\\
         \hline
         28366, 5s & Chen, 2021  & 96.6\% & 69.7\% & 91.3\% & 94.7\%\\
        & Proposed & 99.3\% & 94.5\% & 98.3\% & 98.9\%\\
       \hline\hline
        & \multicolumn{5}{l}{USING QUEENSLAND DATASET- STFT images}\\
         \hline
        Dataset & Author  & Se & Sp & ACC & F1-Score\\
         \hline
        28366, 5s & Chen, 2021  & 96.8\% & 81.5\% & 93.6\% & 96\%\\
       & Proposed & 96.9\% & 83.1\% & 94\% & 96.2\%\\
        \hline
        \end{tabular}
    \label{tab:my_label4}
\end{table}
Apart from comparing with different baseline models [15], the proposed work is compared with different existing works. The dataset used in this paper is the largest one among comparable ones in Table V, and it is 2.5 times larger than the dataset used by earlier DL based work [15].Table V presents such a comparison. Eight existing works were selected and compared with the proposed work regarding accuracy, sensitivity and specificity [5-15].

\begin{table}[h]
    \caption{PERFORMANCE COMPARISON WITH OTHER WORKS}
    \centering
    \setlength\tabcolsep{3 pt}
    \setlength{\extrarowheight}{1pt}
    \begin{tabular}{l cccccc}
    \hline\hline
        Author & Dataset & Subjects & Se & Sp & ACC \\
        \hline 
        Orpahandiou,2015 [3] & 1500,10s & 7 &91\% & 95\% & - \\
        \hline
         Sukor, 2011 [5] & 104, 60s & 13  & 89\% & 77\% & 83\% \\
         \hline
         Li,2012 [6] & 1055, 6s & 104 & 99\% & 80.6\% & 95.2\% \\
         \hline
         Couceiro,2014 [7] & - &  15 & 84.3\% &91.5\% & 88.5\% \\
         \hline
         Cherif,2016 [8] & 104, 6s & - & 84\% & 83\% & 83\% \\
          \hline
         Liu,2020 [9] & 12876, 7s &20 & 91.8\% & 87.3\% & 89.9\% \\
         \hline
         Roh, 2021 [10] & 49561,- & 76 & 96.4\% & \textbf{98.7\%} & 97.5\% \\
         \hline
         Chen,2021 [15] & 5804, 10s & 102 & 98.9\% & 96.7\% & \textbf {98.3\%} \\
         \hline
         Proposed & \textbf{28366, 5s} & 32 & \textbf{99.3\%} & 94.5\% & \textbf{98.3\%} 
         \\
         \hline
    \end{tabular}
    \label{tab:my_label5}
\end{table}
\vspace{-1mm}
\section {CONCLUSION}

As per authors best understanding, this is the first paper to use lightweight deep learning methods combined with a semi classical signal processing based approach for Signal quality assessment of PPG signals. The lightweight novel slim-CNN model takes QPR processed images as input and gives good or bad quality results for PPG signals. The proposed method is validated on a dataset containing PPG data from 32 persons. The novel slim-CNN based model coupled with QPR images outperformed most of the available methods in terms of accuracy, specificity, selectivity, F1-score. The performance of the proposed algorithm is excellent even in the case of the noisy welltory dataset. Results prove that the proposed algorithm can be helpful in the reliable estimation of physiological parameters from PPG signals.


\addtolength{\textheight}{-12cm}   




\end{document}